\documentclass[12pt]{iopart}
\usepackage{graphicx,color,amssymb,amsfonts,hyperref}
\newcommand{\beq}{\begin{equation}}
\newcommand{\eeq}{\end{equation}}
\newcommand{\bqa}{\begin{eqnarray}}
\newcommand{\eqa}{\end{eqnarray}}
\newcommand{\nn}{\nonumber}

\newcommand{\dbd}[1]{\frac{\partial}{\partial {#1}}}
\newcommand{\rt}[1]{\sqrt{#1}\,}
\newcommand{\erf}[1]{Eq.~(\ref{#1})}
\newcommand{\erfs}[1]{Eqs.~(\ref{#1})}

\newcommand{\bra}[1]{\left\langle{#1}\right|}
\newcommand{\ket}[1]{\left|{#1}\right\rangle}

\newcommand{\sq}[1]{\left[ {#1} \right]}
\newcommand{\cu}[1]{\left\{ {#1} \right\}}
\newcommand{\ro}[1]{\left( {#1} \right)}
\newcommand{\an}[1]{\langle{#1}\rangle}

\newcommand{\st}[1]{\left| {#1} \right|}

\renewcommand{\d}{{\rm d}}

\newcommand{\red}[1]{\color{red}{#1}\ \color{black}}

\newtheorem{theorem}{Theorem}

\newtheorem{exercise}[theorem]{Exercise}

\begin{document}

\title[]{Reply to Els\"{a}sser's Comment on ``How many principles does it take to change a light bulb \ldots\ into a laser?''}
\author{Howard M. Wiseman}
\address{
Centre for Quantum Dynamics, Griffith University, Brisbane, Queensland 4111, Australia}
\date{15th September 2015}

\begin{abstract}
In his Comment, Els\"{a}sser claims that the answer to my titular question is one, not four as I have it. He goes 
on to give the singular principle that supposedly captures the difference between a light-bulb and a laser: $g^{(2)}(\tau\!=\!0)=1$. 
His claim is unconsidered and wrong; his proposed principle is impossible to apply and, when corrected, 
redundant (it then becomes one of the four I list already); his arguments are manifestly misdirected. My paper stands as is. 
\end{abstract}

\maketitle

\section{My paper} 
My paper~\cite{Wis16a} addressed the question: ``what are the fundamental features that distinguish laser light from thermal light?'' By the latter I mean the light emitted by a hot body, such as an old-fashioned (incandescent) light bulb. 
This is apparent from my title and also clearly explained in my paper. I identified four necessary fundamental features 
(here in qualitative form): 
\begin{enumerate}
\item High directionality;
\item Monochromaticity;
\item High brightness;
\item Stable intensity.
\end{enumerate}
I explained how to quantify these features and considered these typical devices: a single-mode laser with power of 100 mW, a linewidth $\Gamma$ of $10^7$ s$^{-1}$, and a wavelength $\lambda_0$ of 1 micron; and a light bulb with filament area $A=15$ mm$^2$, and a peak-spectral wavelength $\lambda_{\rm max}$ of 1 micron. The comparison was as follows:
\begin{enumerate}
\item Laser light is polarized and has a single transverse mode; light-bulb light is unpolarized and is emitted into something like $A/\lambda_{\rm max}^2 \sim 10^7$ transverse modes. 
\item Laser light is monochromatic, with $\Gamma/\omega_0 \sim 10^{-8}$; light-bulb light is broad-spectrum with $\delta\omega \sim \omega_{\rm max}$. 
\item Laser light is intense, with ${\nu} \sim  10^{12}$ photons per coherence time; light-bulb light has $\bar{n}_{\rm th}(\omega_{\rm max}) \approx 0.2$ photons per spatio-temporal mode at spectral peak.   
\item Laser light has a stable intensity, with photocount uncertainty 
$\delta N \ll \bar{N}$ over time intervals long enough that $\bar{N} \gg 1$; 
light-bulb light, if collimated and filtered, would have a photocount uncertainty $\delta N$ larger than $\bar{N}$ for time intervals $\lesssim \Gamma^{-1}$. 
\end{enumerate} 

\section{Els\"{a}sser's claim and principle}

Els\"{a}sser~\cite{Elsasser}, in contrast to me, claims that only one principle is necessary to ``understand the difference between \ldots\ a light bulb and a laser'': 
\begin{quote}
thermal light showing a central second order correlation $g^{(2)}(\tau\!=\!0)$ value  of two is confronted with that 
of a value of unity for laser light.
\end{quote}
This claim and principle may be ``simple, elegant, clear and unique'', as he says, but it is also unconsidered, wrong, inapplicable, and redundant 
(when made applicable). 

Els\"{a}sser's claim is {\em unconsidered} in that his principle presupposes the existence of a beam so that the 
second-order correlation function for photon-counting $g^{(2)}$ is a function of only a single co\"ordinate 
(which he labels as $\tau$). That is, Els\"{a}sser has implicitly assumed my first principle even while 
claiming it to be unnecessary. 

Els\"{a}sser's  claim is {\em wrong} in that it is simple to imagine a type of light that satisfies his principle 
$g^{(2)}(\tau\!=\!0)=1$ but which could not possibly be considered laser light. As Els\"{a}sser notes, non-classical 
light can exhibit $g^{(2)}(0)<1$. 
An example of a system that can achieve $g^{(2)}(0)<1$ is the radiation 
emitted by a single atom (considered at a single point in the radiated field, for example). Consider a detector 
behind a pin-hole in a screen close to such an atom. If there is, on the same side of the screen as the atom, 
but a long way away, a light bulb, then the $g^{(2)}(0)=2$ thermal light will add incoherently to the 
$g^{(2)}(0)<1$ atomic fluorescence. Clearly this sum could be made as close as desired to $g^{(2)}(0)=1$, 
but that doesn't mean that suddenly one has made a laser beam! 

Els\"{a}sser's principle is {\em inapplicable} in that $g^{(2)}(\tau\!=\!0)=1$ is an unachievable ideal. 
Indeed he goes on to say that a laser has Poissonian statistics, which implies the even stronger condition $g^{(2)}(\tau)=1$ 
for all $\tau$. This is certainly not true for real lasers, 
as is manifest in the 
fact that at low frequencies
technical noise is far larger than the Poissonian 
or shot noise of the ideal laser Els\"{a}sser imagines. 
We could replace Els\"{a}sser's inapplicable $g^{(2)}(\tau)=1$ by $g^{(2)}(\tau)\approx 1$ --- 
that is, $|g^{(2)}(\tau)-1| \ll 1$ ---
which can still be true even with large technical noise.  
However $g^{(2)}(\tau)$ could, in principle, take values quite different from one only on a set of 
arbitrarily small measure on the real line and this would have arbitrarily small observable consequences. 
A less stringent requirement is that a suitable {\em average} 
of $g^{(2)}(\tau)$ is very close to one over {\em any} interval $[0,T]$  such that $\bar{I}T \gg 1$. 
The physical significance of this condition is as follows. The 
uncertainty $\delta N$ in the number of photons in an interval $[0,T]$ is given by 
\bqa
 (\delta N)^2 &=& \int_0^{T} ds \int_0^{T} dt \cu{\bar{I}^2[g^{(2)}(s-t)-1] + \bar{I}\delta(s-t)}  \\
 &=& \bar{N} + \bar{N}^2  [\bar{g}_T^{(2)}-1],
\eqa
 where $\bar{N} = \bar{I}T \gg 1$ is the mean number of photons in that interval and 
 $\bar{g}_T = T^{-2} \int_0^{T} dT' \int_{0}^{T'} d\tau  g(\tau)$. Thus, if and only if 
 $\bar{g}_T \approx 1$, we have $\delta N \ll \bar{N}$.

 
Els\"{a}sser's principle, when made physically applicable in the above way, is thus {\em redundant}. 
The condition that this average of 
$g^{(2)}(\tau)$, over any interval large enough to contain a macroscopic field (many photons), be close to unity, 
is implied by my condition (iv), that the laser have a stable intensity.  

\section{Els\"{a}sser's subsequent arguments} 

Following the above, Els\"{a}sser seems to take aim at my principles (ii) and (iii), saying (my emphasis added)
\begin{quote}
{\em only considering} the spectral properties in terms of the first order (field) correlation $g^{(1)}(\tau\!=\!0)$ 
obtained via the Wiener-Khintchine theorem is no longer a criterion for differentiating between thermal
and laser light.
\end{quote}
If this is his aim, he misses wildly. First, $g^{(1)}(0)=1$ by definition, for any field, so it can not be used as a criterion for anything. Second, while my principles (ii) and (iii) do involve the first-order correlation 
$g^{(1)}(\tau)$, the latter principle cannot be stated in terms of this correlation function alone. Third, 
it could not be more explicit in my paper that I characterise laser light by a {\em conjunction} of four features. I am never ``only considering'' one of my principles. 

Similarly, when Els\"{a}sser turns at last to my principle (i), saying 
\begin{quote}
directionality is no longer a unique criterion for laser light because amplified spontaneous emission
originating from semiconductor-based optoelectronic light emitters with waveguides unifies broad-band and
directionality and does exhibit photon bunching, i.e.~thermal light second order coherence characteristics.
\end{quote}
he seems to have forgotten the title of his own paper (quoting, as it does, the title of mine). The question is not, 
``how many principles does it take to change {\em amplified spontaneous emission
originating from semiconductor-based optoelectronic light emitters with waveguides} into a laser?'' It is 
``how many principles does it take to change a {\em light bulb} into a laser?'' The fact that  
amplified spontaneous emission originating from semiconductor-based optoelectronic light emitters with waveguides 
has ``thermal light second order coherence characteristics'' does not mean that it is 
thermal light, like that from a light bulb. Glauber~\cite{Gla63} used the term {\em incoherent light} to refer to 
light with the same intensity correlations as thermal light, regardless of its other properties,  
as I discuss in my Conclusion. The whole of my Section 4 addresses the point that it is 
possible to have light that satisfies my first three principles, while remaining incoherent 
in terms of its intensity correlations. Els\"{a}sser is attacking a straw man. 

\section{Discussion}

Els\"{a}sser's fixation on $g^{(2)}(0)=1$ as {\em the} defining feature of a laser is certainly not supported 
by scientists who have sought to communicate the importance of the laser to the public. As I quoted in my paper,  
the official {\em Year of Light} (2015) home page~\cite{iyolLaser} says (my emphasis added).
\begin{quote}
A laser is an optical amplifier --- a device that strengthens light waves.  Some lasers have a {\em well-directed}, {\em very bright} beam with a {\em very specific color}\,; others emphasize different properties, such as extremely short pulses.  The key feature is that the amplification makes light that is very {\em well defined and reproducible}, unlike ordinary light sources such as the {\em sun or a lamp}.
\end{quote}
The sources with which a laser is contrasted are thermal (in all respects).  
The first three features I list appear prominently\footnote{It is true that they are listed as properties only of ``some lasers'', 
and the other example given, of a laser producing extremely short pulses, is clearly not monochromatic, the qualitative statement of my criterion (ii). 
However, the most useful lasers producing extremely short pulses have a locked carrier-envelope phase~\cite{CEPlocking2000}, and would  
still satisfy my quantitative criterion (ii), $\Gamma \ll \omega_0$,  
if the linewidth $\Gamma$ is taken to be the width of the narrow peaks within the broad spectrum, 
rather than the width of the broad spectrum itself.}.
The fourth feature,  
a stable intensity, is referenced obliquely, at best, if it is covered by the phrase ``well defined and reproducible''.

To conclude, Els\"{a}sser's comment has no bearing on the correctness, relevance, or cogency of my paper. 
I hope that, in time, it will repay his closer attention. 

\section*{References}
\bibliography{How_many_principles-reply.bbl}

\end{document}